%% file: automatica_brief.tex
\documentclass[twocolumn]{autart}    

\usepackage{graphicx}          

\usepackage{amsmath} 
\usepackage{amssymb} 
\usepackage{tikz}
\usepackage{xcolor}
\usepackage{url}
\usepackage{comment}

\newtheorem{lemma}[thm]{Lemma}
\newtheorem{proposition}[thm]{Proposition}
\newtheorem{theorem}[thm]{Theorem}

\begin{document}

\begin{frontmatter}

\title{Distributed formation maneuver control by manipulating the complex Laplacian} 


\author[Spain]{Hector Garcia de Marina}\ead{hgarciad@ucm.es}    

\address[Spain]{Department of Computer Architecture and Automatic Control. Faculty of Physics. Universidad Complutense Madrid.}  


\begin{abstract}                          
This paper proposes a novel maneuvering technique for the \emph{complex-Laplacian-based} formation control. We show how to modify the original weights that build the Laplacian such that a designed steady-state motion of the desired shape emerges from the local interactions among the agents. These collective motions can be exploited to solve problems such as the \emph{shaped consensus} (the rendezvous with a particular shape), the \emph{enclosing of a target}, or \emph{translations with controlled speed and heading} to assist mobile robots in area coverage, escorting, and traveling missions, respectively. The designed steady-state collective motions correspond to rotations around the centroid, translations, and scalings of a reference shape. The proposed modification of the weights relocates one of the Laplacian's zero eigenvalues while preserving its associated eigenvector that constructs the desired shape. For example, such relocation on the imaginary or real axis induces rotational and scaling motions, respectively. We will show how to satisfy a sufficient condition to guarantee the global convergence to the desired shape and motions. Finally, we provide simulations and comparisons with other maneuvering techniques.
\end{abstract}

\end{frontmatter}

\input{paper_brief.tex}

\begin{ack}                               
This work has been supported by the grant \emph{Atraccion de Talento} with reference number 2019-T2/TIC-13503 cofunded by the Government of the Autonomous Community of Madrid, and it has been partially supported by the Spanish Ministry of Science and Innovation under research Grant RTI2018-098962-B-C21.
\end{ack}

\bibliographystyle{plain}        
\bibliography{Bibs}              



\end{document}

%% file: paper_brief.tex

\section{Introduction}
The scientific community and industry anticipate distributed robot swarms assisting humans in challenges involving vast, hard accessible, and dangerous areas \cite{yang2018grand}. These challenges are related to specific missions in environmental monitoring, intensive agriculture, search \& rescue, and disaster management, among others, \cite{delmerico2019current}. In particular, the \emph{distributed} nature of these groups of robots (or agents in general) concentrates on the local interaction between the individuals to create collective behaviors. One of these global behaviors focuses on displaying geometrical patterns to assist the team in higher-level tasks \cite{Oh2015}. In this regard, formation control algorithms offer a repertoire of solutions depending on the sensing capabilities of the agents and the desired geometrical pattern. Far from being a solved problem, scientists are still on the development of reliable methods for the control and coordination of robot swarms, or multi-agent systems in general \cite{yang2018grand}. In robotics, it is common to demand from the swarm not only to display a shape but to move in a coordinated fashion.

In this paper, we focus on maneuvering formations of multi-agent systems based on the complex Laplacian matrix \cite{lin2014distributed}. In particular, we show that it is possible to achieve the (simultaneous) coordinated motions of translation, rotation, and scaling by only modifying a set of the original complex weights, designed only for static formations, in the Laplacian matrix. This fact enables us to preserve three interesting properties. Firstly, it is \emph{distributed}, i.e., an individual agent only needs local information. 
Secondly, the agents do not need to share any common frame of coordinates. Thirdly, the agents control \emph{tensions}, e.g., they aim to have the zero weighted sum of their available relative positions for having an eventual static formation. We will show that a designed non-zero weighted sum of the relative positions will be responsible for the collective motion. Indeed, that is why we propose the modification of the original weights in \cite{lin2014distributed}. 

The simplicity of our maneuvering technique on modifying the weights can be analyzed in detail by explicitly solving the resultant matrix differential equation involving the \emph{modified} complex Laplacian matrix. We prove how to \emph{relocate} one of the two Laplacian's original zero eigenvalues while we keep its associated eigenvector, which describes the \emph{appearance} of the desired shape. We move such a zero eigenvalue over the real and the imaginary axis to achieve scaling and rotational motions. For pure translations, we prove that our technique reduces from two to one the geometric multiplicity of the zero eigenvalue. The relocation of such zero eigenvalue of the original complex Laplacian matrix comes with an inconvenience, i.e., the rest of the non-zero eigenvalues are relocated as well. We provide an explicit condition on a constant that can be satisfied by design such that the original non-zero eigenvalues do not cross the imaginary axis, i.e., we can guarantee the global convergence to the desired steady-state motion and shape simultaneously.

We present three practical applications as the motivation for the proposed maneuvering technique, namely, the \emph{shaped consensus}, the \emph{enclosing of a target}, and \emph{the travelling formation with controlled speed and heading}. Shaped consensus consists in the rendezvous of all the agents while they display the desired shape, in contrast with the \emph{standard} consensus algorithm where the shape is not under control \cite{olfati2007consensus}. Also, this shaped consensus can be done while describing an inwards or outwards circular spiral trajectory. Both collective motions (inwards and outwards) can be of interest in area coverage scenarios. The enclosing of a target maneuver chooses one agent and has the rest orbiting around it with a constant angular speed. For example, this collective motion is of interest in escorting missions. In contrast with other approaches \cite{lan2010distributed,Mar12}, among other works, we do not require \emph{all} the enclosing agents to track the target, or to follow the same circular path. Finally, we will show that the formation can travel with an arbitrary speed and heading controlled by only one agent, i.e., without the need of any \emph{leader-follower} architecture/estimators as it is common in the literature \cite{lin2013leader,zhao2018affine}.


This paper has been organized in the following way. Section \ref{sec: pre} introduces the required notation and the notion of \emph{desired shape}. In Section \ref{sec: dyn}, we debrief the complex-Laplacian-based formation control, and we introduce our strategy on modifying the Laplacian's weights to induce collective motions in Section \ref{sec: modL}. In Section \ref{sec: complex}, we derive the solutions of the resulting linear system explicitly after applying our maneuvering technique; in particular, we show how to satisfy a sufficient condition such that the global convergence to the desired collective motion is guaranteed. We continue in Section \ref{sec: simulations}, discussing the applications of our technique with illustrative simulations. Finally, we end the paper in Section \ref{sec: conc} with some conclusions and future work.

\section{Preliminaries}
\label{sec: pre}
\subsection{Notation and graph theory}
We consider the complex-Laplacian-based formation control of $n \in \mathbb{N} \geq 2$ mobile agents on the plane. We represent the complex unit by the symbol $\iota$. We denote by $||x||$ the Euclidean norm of the vector $x\in\mathbb{C}^p, p\in\mathbb{N}$. Given a set $\mathcal{X}$, we denote by $|\mathcal{X}|$ its cardinality. Finally, we denote by $\mathbf{1}_p\in\mathbb{C}^p, p\in\mathbb{N}$, the all-one column vector.

A \emph{graph} $\mathcal{G} = (\mathcal{V}, \mathcal{E})$ consists of two non-empty sets: the node set $\mathcal{V} = \{1,2,\dots,n\}$, and the edge set $\mathcal{E} \subseteq (\mathcal{V}\times\mathcal{V})$. In this paper we deal with the special case of \emph{undirected} graphs. In particular, undirected graphs are \emph{bidirectional} graphs where if the edge $(i,j)\in\mathcal{E}$, then the edge $(j,i)\in\mathcal{E}$ as well. The set $\mathcal{N}_i$ containing the neighbors of the node $i$ is defined by $\mathcal{N}_i:=\{j\in\mathcal{V}:(i,j)\in\mathcal{E}\}$. Let $w_{ij}\in\mathbb{C} \neq 0$ be a weight associated with the edge $(i,j)\in\mathcal{E}$, then the complex \emph{Laplacian} matrix $L\in\mathbb{C}^{n\times n}$ of $\mathcal{G}$ is defined as
\begin{equation}
	l_{ij} := \begin{cases}\sum_{k\in\mathcal{N}_i}w_{ik} & \text{if} \quad i = j \\
		-w_{ij} & \text{if} \quad i \neq j \wedge j\in\mathcal{N}_i \\
		0 & \text{if} \quad i \neq j \wedge j\notin\mathcal{N}_i.
	\end{cases}
	\label{eq: L}
\end{equation}
We note that if $\mathcal{G}$ is connected, then $L\mathbf{1}_n = 0$. For an undirected graph, we choose one of the two arbitrary directions for each pair of neighboring nodes to construct the ordered set of edges $\mathcal{Z}$. For an arbitrary edge $\mathcal{Z}_k = (\mathcal{Z}_k^{\text{head}},\mathcal{Z}_k^{\text{tail}}), k\in\{1,\dots,\frac{|\mathcal{E}|}{2}\}$, we call to its first and second element the \emph{tail} and the \emph{head} respectively. From such an ordered set, we construct the following \emph{incidence matrix} $B\in\mathbb{C}^{|\mathcal{V}|\times |\mathcal{Z}|}$ that satisfies $B^T\mathbf{1}_n = 0$:
\begin{equation}
	b_{ik} := \begin{cases}+1 \quad \text{if} \quad i = {\mathcal{Z}}^{\text{tail}}_k \\
		-1 \quad \text{if} \quad i = \mathcal{Z}^\text{head}_k \\
		0 \quad \text{otherwise.}
	\end{cases}
	\label{eq: B}
\end{equation}

\subsection{Frameworks and desired shape}
We codify the 2D position of each agent $i\in\mathcal{V}$ in $p_i\in\mathbb{C}$, where we take the real and imaginary parts respectively for the coordinates of the two dimensions of the Euclidean plane. We stack all the positions $p_i$ in a single vector $p\in\mathbb{C}^{n}$ and we call it \emph{configuration}. We define a \emph{framework} $\mathcal{F}$ as the pair $(\mathcal{G}, p)$, where we assign each agent's position $p_i$ to the node $i\in\mathcal{V}$, and the graph $\mathcal{G}$ establishes the set of neighbors $\mathcal{N}_i$ for each agent $i$.

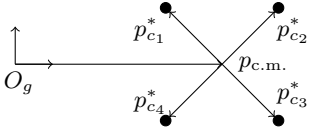
\begin{figure}
\centering
\begin{tikzpicture}[line join=round]
\filldraw(0,0) circle (2pt);
\filldraw(1.5,0) circle (2pt);
\filldraw(1.5,1.5) circle (2pt);
\filldraw(0,1.5) circle (2pt);
\draw[draw=black,arrows=->](-2,0.75)--(-1.5,0.75);
\draw[draw=black,arrows=->](-2,0.75)--(-2,1.25);
\draw[draw=black,arrows=->](-2,0.75)--(0.75,0.75);
\draw[draw=black,arrows=->](0.75,0.75)--(0+0.05,1.5-0.05);
\draw[draw=black,arrows=->](0.75,0.75)--(0+0.05,0+0.05);
\draw[draw=black,arrows=->](0.75,0.75)--(1.5-0.05,1.5-0.05);
\draw[draw=black,arrows=->](0.75,0.75)--(1.5-0.05,0+0.05);
\node at (-1.95,0.5) {\small $O_g$ \normalsize};
\node at (0-0.2,1.2) {\small $p_{c_1}^*$ \normalsize}; 
\node at (0-0.2,0.75-0.5) {\small $p_{c_4}^*$ \normalsize}; 
\node at (1.5+0.2,0.75+0.45) {\small $p_{c_2}^*$ \normalsize}; 
\node at (1.5+0.2,0.75-0.4) {\small $p_{c_3}^*$ \normalsize}; 
\node at (1.3,0.75) {\small $p_{\text{c.m.}}$ \normalsize}; 
\end{tikzpicture}
	\caption{The vector $p^* =  p_{\text{c.m.}}\mathbf{1}_4 + [p_{c_1}^* \, \dots \, p_{c_4}^*]^T$ is an example of \emph{reference shape} to construct a \emph{desired shape} $\mathcal{S}$.}
\label{fig: pstar}
\end{figure}

We choose an arbitrary configuration of interest or \emph{reference shape $p^*$} for the team of agents, and we split it as
\begin{equation}
	p^* = p_{\text{c.m.}}\mathbf{1}_n  + p^*_c,
	\label{eq: pstar}
\end{equation}
where $p_{\text{c.m.}}\in\mathbb{C}$ is the position of the \emph{center of mass} of the configuration and $p_c^*\in\mathbb{C}$, starting from $p_{\text{c.m.}}$, gives the \emph{appearance} to the formation as in the example shown in Figure \ref{fig: pstar}. Without loss of generality, and for the sake of simplicity, we set $p_{\text{c.m.}} = 0$ in (\ref{eq: pstar}), i.e., $p^* = p^*_c$.

We now define the concept of \emph{desired shape} constructed from the reference shape $p^*$:
\begin{defn}
The framework or the formation is at the \emph{desired shape} when
	\begin{equation}
		p\in\mathcal{S}:=\{p = c_1 \mathbf{1}_{n} + c_2 p^* \, | \, c_1,c_2\in\mathbb{C}\},
	\label{eq: dshape}
\end{equation}
\end{defn}
Note that $c_1$ accounts for translations without restrictions in the space, while $c_2$ accounts for the scaling and rotation of all the elements of the reference $p^*$ equally.


\section{Agents' dynamics and shape stabilization}
\label{sec: dyn}
We consider that the position of each agent is modelled by the single-integrator dynamics
\begin{equation}
	\dot p_i = u_i, \quad \forall i\in\mathcal{V},
	\label{eq: dyn}
\end{equation}
where $u_i\in\mathbb{C}$ is the control action for the corresponding agent $i$. 
Later, for the analysis of the whole system, let us consider the following compact form for all the agent's dynamics in (\ref{eq: dyn})
\begin{equation}
\dot p = u,
	\label{eq: pdyn}
\end{equation}
where $u\in\mathbb{C}^{n}$ is the stacked vector of control actions $u_i$.

We will consider only distributed control actions; therefore the agent $i$ only has access to relative information with respect to its neighbors in $\mathcal{N}_i$. In particular, such a local available information is the set of relative positions $z_{ij} = p_i - p_j, (i,j)\in\mathcal{E}$. 

In order to drive $p(t)$ to $\mathcal{S}$, our technique will start from the distributed complex-Laplacian-based formation control analyzed in \cite{lin2014distributed}, i.e., 
\begin{equation}
	u_i = -hk_i \sum_{j\in\mathcal{N}_i} \omega_{ij} (p_i - p_j) = -hk_i\sum_{j\in\mathcal{N}_i} \omega_{ij} z_{ij},
	\label{eq: uLcomplex}
\end{equation}
where $h\in\mathbb{R}^+$ is an arbitrary positive gain, $\omega_{ij} \in \mathbb{C}$ is a complex weight to be designed later for the construction of the complex Laplacian matrix (\ref{eq: L}), and $k_i\in\mathbb{C} \setminus \{0\}$ is a non-zero gain to be designed to guarantee the convergence of the formation to the desired (static) shape. Considering all the control actions (\ref{eq: uLcomplex}), we can arrive at the following closed loop in compact form
\begin{equation}
	\dot p = -hKLp,
	\label{eq: comcom}
\end{equation}
where $K = \operatorname{diag}\{k_1, \dots, k_n\}$. We need the assistance of the matrix $K$ in (\ref{eq: comcom}) since some of the non-zero eigenvalues of the complex $L$ might be in the left-half plane. Note that $L$ is not positive semi-definite necessarily since $\omega_{ij}\neq\omega_{ji}$ in general. For regular polygonal formations with $(n-1)$ edges, it has been proved that $K=I_n$ is sufficient to guarantee the stability of (\ref{eq: comcom}) \cite{Marina2017}.


The authors in \cite{lin2014distributed} identified a (graph) requirement for the design of the weights in (\ref{eq: uLcomplex}). In particular, we can see that the desired shape must satisfy the following $n$ linear constraints
\begin{equation}
	\sum_{j\in\mathcal{N}_i} \omega_{ij} (p_{i}^* - p_{j}^*) = 0,\quad \forall i\in\mathcal{V}.
	\label{eq: wcomcon}
\end{equation}
These conditions can be satisfied if and only if the graph $\mathcal{G}$ is \emph{2-rooted}, i.e., if there exists a subset of two nodes, from which every other node is \emph{2-reachable} \cite{lin2014distributed}. A node $v\in\mathcal{V}$ is \emph{2-reachable} from a non-singleton set $\mathcal{U}$ of nodes if there exists a path from a node in $\mathcal{U}$ to $v$ after removing any one node except node $v$.


The complex Laplacian, with its weights satisfying the constrains (\ref{eq: wcomcon}), has two zero eigenvalues with eigenvectors $\mathbf{1}_n$ and $p^*$ respectively. Therefore, if the rest of eigenvalues of $KL$ are in the right half plane, then $p(t)$ in (\ref{eq: comcom}) converges to the kernel of the complex $L$, i.e., $p(t) \to \mathcal{S}$ as $t\to\infty$. In fact, the configuration $p(t)$ converges to a point in $\mathcal{S}$, i.e., the formation stops eventually. 

\section{Modified Laplacian matrix}
\label{sec: modL}
The maneuvering technique in this paper consists in modifying a (non-unique) subset of the weights $\omega_{ij}$ in (\ref{eq: wcomcon}) such that the formation converges to a steady-state motion within the desired shape $\mathcal{S}$. In particular, the modification of the weights $\omega_{ij}$ will be designed by exploiting the available (given by $\mathcal{E}$) relative positions between the agents in the reference shape $p^*$.

Let us consider the following weights to construct a modified Laplacian matrix
\begin{equation}
	\tilde\omega_{ij} = \omega_{ij} - \frac{\tilde\kappa}{hk_i}\mu_{ij}, \quad (i,j)\in\mathcal{E}
\label{eq: wmod}
\end{equation}
where the \emph{motion parameters} $\mu_{ij}\in\mathbb{C}$ will be designed in Section \ref{sec: mpd} for the translation, rotation and scaling of the formation, and $\tilde\kappa\in\mathbb{R}$ will regulate the speed and direction of such motions. We recall that $k_i$ and $h$ come from the gains in (\ref{eq: uLcomplex}), and we will need them to compensate for $hK$ in the design of the steady-state motions. Since our maneuvering technique is also distributed, then if $j\notin\mathcal{N}_i$, then $\mu_{ij} = 0$. In general, we will have that $\mu_{ij} \neq \mu_{ji}$ and $\tilde\omega_{ij} \neq \tilde\omega_{ji}, (i,j)\in\mathcal{E}$, for the modified Laplacian matrix.

Similarly to the incidence matrix $B$ in (\ref{eq: B}), consider again the ordered set of edges $\mathcal{Z}$, and let us define the components of the following matrix $M\in\mathbb{C}^{|\mathcal{V}|\times |\mathcal{Z}|}$
\begin{equation}
	m_{ik} := \begin{cases}\mu_{i\mathcal{Z}^\text{head}_k} \quad \text{if} \quad i = \mathcal{Z}^\text{tail}_k \\
		-\mu_{i\mathcal{Z}^\text{tail}_k} \quad \text{if} \quad i = \mathcal{Z}^\text{head}_k \\
		0 \quad \text{otherwise.}
	\end{cases}.
	\label{eq: M}
\end{equation}
The definition (\ref{eq: M}) enables us to write the modified Laplacian matrix from the modified weights (\ref{eq: wmod}) in compact form as
\begin{equation}
	\tilde L = L - \frac{\tilde\kappa}{h} K^{-1}MB^T.
	\label{eq: Ltilde}
\end{equation}

\section{Shape maneuvering}
\label{sec: complex}

\subsection{Motion parameters design}
\label{sec: mpd}

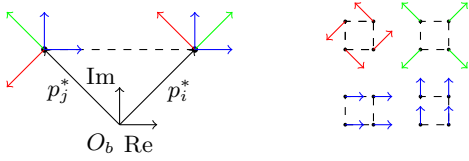
\begin{figure}
\centering
\begin{tikzpicture}[line join=round]
\begin{scope}[shift={(-4,0)}]
\filldraw(1,1) circle (1pt);
\filldraw(-1,1) circle (1pt);
\draw[dashed](1,1)--(-1,1);
\draw[draw=black,arrows=->](0,0)--(0.5,0);
\draw[draw=black,arrows=->](0,0)--(0,0.5);
\draw[draw=black,arrows=->](0,0)--(1,1);
\draw[draw=black,arrows=->](0,0)--(-1,1);
\draw[draw=black,color=green,arrows=->](1,1)--(1.5,1.5);
\draw[draw=black,color=red,arrows=->](1,1)--(0.5,1.5);
\draw[draw=black,color=blue,arrows=->](1,1)--(1.5,1);
\draw[draw=black,color=blue,arrows=->](1,1)--(1,1.5);
\draw[draw=black,color=green,arrows=->](-1,1)--(-1.5,1.5);
\draw[draw=black,color=red,arrows=->](-1,1)--(-1.5,0.5);
\draw[draw=black,color=blue,arrows=->](-1,1)--(-0.5,1);
\draw[draw=black,color=blue,arrows=->](-1,1)--(-1,1.5);
	\node at (-0.25,-0.25) {\small $O_b$ \normalsize};
	\node at (0.25,-0.25) {\small Re \normalsize};
	\node at (-0.25,0.65) {\small Im \normalsize};
\node at (0.8, 0.45) {\small $p_{i}^*$ \normalsize};
\node at (-1+0.2,0.45) {\small $p_{j}^*$ \normalsize};
\end{scope}
\begin{scope}[shift={(-1,0)},scale=0.25]
\filldraw(0,0) circle (2pt);
\filldraw(1.5,0) circle (2pt);
\filldraw(1.5,1.5) circle (2pt);
\filldraw(0,1.5) circle (2pt);
\draw[draw=black,dashed](0,0)--(1.5,0)--(1.5,1.5)--(0,1.5)--(0,0);
\draw[draw=black,color=blue,arrows=->](0,0)--(1,0);
\draw[draw=black,color=blue,arrows=->](1.5,0)--(2.5,0);
\draw[draw=black,color=blue,arrows=->](1.5,1.5)--(2.5,1.5);
\draw[draw=black,color=blue,arrows=->](0,1.5)--(1,1.5);
\end{scope}
\begin{scope}[shift={(0,0)},scale=0.25]
\filldraw(0,0) circle (2pt);
\filldraw(1.5,0) circle (2pt);
\filldraw(1.5,1.5) circle (2pt);
\filldraw(0,1.5) circle (2pt);
\draw[draw=black,dashed](0,0)--(1.5,0)--(1.5,1.5)--(0,1.5)--(0,0);
\draw[draw=black,color=blue,arrows=->](0,0)--(0,1);
\draw[draw=black,color=blue,arrows=->](1.5,0)--(1.5,1);
\draw[draw=black,color=blue,arrows=->](1.5,1.5)--(1.5,2.5);
\draw[draw=black,color=blue,arrows=->](0,1.5)--(0,2.5);
\end{scope}
\begin{scope}[shift={(-1,1)},scale=0.25]
\filldraw(0,0) circle (2pt);
\filldraw(1.5,0) circle (2pt);
\filldraw(1.5,1.5) circle (2pt);
\filldraw(0,1.5) circle (2pt);
\draw[draw=black,dashed](0,0)--(1.5,0)--(1.5,1.5)--(0,1.5)--(0,0);
\draw[draw=black,dashed](0,0)--(1.5,0)--(1.5,1.5)--(0,1.5)--(0,0);
\draw[draw=black,color=red,arrows=->](0,0)--(1,-1);
\draw[draw=black,color=red,arrows=->](1.5,0)--(2.5,1);
\draw[draw=black,color=red,arrows=->](1.5,1.5)--(0.5,2.5);
\draw[draw=black,color=red,arrows=->](0,1.5)--(-1,0.5);
\end{scope}
\begin{scope}[shift={(0,1)},scale=0.25]
\filldraw(0,0) circle (2pt);
\filldraw(1.5,0) circle (2pt);
\filldraw(1.5,1.5) circle (2pt);
\filldraw(0,1.5) circle (2pt);
\draw[draw=black,dashed](0,0)--(1.5,0)--(1.5,1.5)--(0,1.5)--(0,0);
\draw[draw=black,color=green,arrows=->](0,0)--(-1,-1);
\draw[draw=black,color=green,arrows=->](1.5,0)--(2.5,-1);
\draw[draw=black,color=green,arrows=->](1.5,1.5)--(2.5,2.5);
\draw[draw=black,color=green,arrows=->](0,1.5)--(-1,2.5);
\end{scope}
\end{tikzpicture}
	\caption{The velocity $^bv_i^*$ for the agent $i$ can be split into four \emph{orthogonal} velocities, as it is shown on the right side for a square formation, such that $p^*$ stays in $\mathcal{S}$. In red color, the velocities for the rotation around $O_b$ (the centroid of the formation). In green color, the scaling velocity must be parallel to the agent's position. In blue color, the two orthogonal translational velocities are arbitrary but equal to all the agents. Note that each of these four velocities (represented as a complex number) can be constructed by multiplying the relative complex number $(p_{i}^* - p_{i}^*)$ (dashed relative position) by an appropriate complex number $\mu_{ij}$.}
\label{fig: velcom}
\end{figure}
Similarly as in ($\ref{eq: wcomcon}$), we will design the parameters $\mu_{ij}$ by satisfying the linear constraint
\begin{equation}
	{^b}v_i^* = \sum_{j\in\mathcal{N}_i}\mu_{ij}\, ({^b}p_i^* - {^b}p_j^*)  = \sum_{j\in\mathcal{N}_i}\mu_{ij}\,{^b}z_{ij}^*, \, \forall i\in\mathcal{V},
\label{eq: vbi}
\end{equation}
where ${^b}v_i^*\in\mathbb{C}$ is the desired velocity for each agent $i$ with respect to a frame of coordinates $O_b$ fixed at the centroid of $p^*$ as shown in Figure \ref{fig: velcom}. Note that in order to satisfy (\ref{eq: vbi}), we need for the agent $i$ to have at least one non-zero $z_{ij}^*$. For example, our technique does not work for $p^* = \mathbf{1}_n$. Since $\mu_{ij}$ \emph{scales and rotates} the complex number ${^b}z_{ij}^*$ (representing a relative position), only one non-zero $\mu_{ij}$ is enough per agent $i$ to construct an arbitrary desired ${^b}v_i^*$. For example, choose an arbitrary neighbor $\tilde j\in\mathcal{N}_i$, and calculate directly
\begin{equation}
	\mu_{ij} := \begin{cases}{^b}v_i^* \, / \, {^b}z_{ij}^* \quad \text{if} \quad j=\tilde j \\
		0 \quad \text{otherwise}
	\end{cases}, \quad \forall i\in\mathcal{N}.
	\label{eq: mus}
\end{equation}

We can stack (\ref{eq: vbi}) for all the agents and arrive at the following compact form
\begin{equation}
	{^b}v_f = MB^T\,{^b}p^*,
	\label{eq: vf}
\end{equation}
where ${^b}v_f \in \mathbb{C}^n$ is the stacked vector with all the desired agents' velocities. In fact, in order to keep a formation invariant in $\mathcal{S}$, we can split ${^b}v_f$ into the four terms as in Figure \ref{fig: velcom}, i.e.,
\begin{equation}
	{^b}v_f = ({^b}v_x^* + {^b}v_y^*)\mathbf{1}_n + a\,{^b}p^* + \iota \omega \, {^b}p^*,
	\label{eq: dvf}
\end{equation}
where $v_{\{x,y\}}^*\in\mathbb{C}$ account for the common (horizontal and vertical) translational velocity in \emph{distance units}/sec, $a\in\mathbb{R}$ sets whether the formation grows or contracts in \emph{current size}/sec, and $\omega\in\mathbb{R}$ sets the angular speed in radians/sec. Likewise, we can split $M$ in (\ref{eq: vf}) into four terms, namely
\begin{equation}
	M = \kappa_{t_x}M_{t_x} + \kappa_{t_y}M_{t_y} +\kappa_s M_s + \kappa_rM_r,
	\label{eq: ormo}
\end{equation}
where the matrices $M_{t_{\{x,y\}}},M_r,M_s\in\mathbb{C}^{|\mathcal{V}|\times|\mathcal{Z}|}$ have their elements $\mu_{ij}$ as in (\ref{eq: M}) but designed only for the $1$ \emph{distance units}/sec translation in horizontal/vertical direction, for the $1$ \emph{current size}/sec scaling, and for the $1$ radian/sec rotation of the reference shape respectively. Finally, we can see $\kappa_{\{t_{\{x,y\}},r,s\}}\in\mathbb{R}$ as the \emph{coordinates} of the four orthogonal motions (see the right side in Figure \ref{fig: velcom}) that will define the eventual collective motion. For example, if $\kappa_{\{r,s\}} = -1$ and $\kappa_{t{\{x,y\}}}=0$, then, the eventual collective motion will be the contraction of the reference shape while its centroid is fixed but the agents spin around it. Note that (\ref{eq: dvf}) and (\ref{eq: ormo}) are linked through (\ref{eq: vf}), and the elements of $M_{t_{\{x,y\}}},M_r$ and $M_s$ have been designed such that $|{^b}v_{\{x,y\}}^*| = |\kappa_{\{t_{\{x,y\}}}|$, $a = \kappa_s$, and $\omega = \kappa_r$.



\subsection{Modified complex Laplacian matrix for translational, rotational and scaling motion control}
The term $\frac{\tilde\kappa}{h}K^{-1}\tilde MB^T$ in (\ref{eq: Ltilde}) modifies the properties of $L$ substantially. In particular, we modify the original two zero eigenvalues of $L$ in different ways while preserving the associated eigenvectors that define the desired shape $\mathcal{S}$. After analyzing such differences, we will see that the formation in the closed-loop with the modified Laplacian
\begin{equation}
	\dot p = -hK\tilde Lp = (-hKL + \tilde\kappa MB^T)p,
	\label{eq: pmov}
\end{equation}
will converge to a moving configuration $p(t)$ in $\mathcal{S}$. We can see $\tilde\kappa$ in (\ref{eq: pmov}) as a gain that regulates the \emph{global speed} of the collective motion once it is defined with $\kappa_{\{t_{\{x,y\}},r,s\}}$ in (\ref{eq: ormo}). Once the \emph{motion gains} are set, then the gain $h$ in (\ref{eq: pmov}) will assist us with having all the original non-zero \emph{stable} eigenvalues of $KL$ no traspassing the imaginary axis in the modified $K\tilde L$. For the following results, let us assume for now that $h$ is large enough so that the second term in (\ref{eq: Ltilde}) is a small enough perturbation of $L$. Later, we will provide a lower bound to $h$ in Subsection \ref{sec: kappa}.


\begin{lemma}
\label{lem: eigtr}
We consider the following two mutually exclusive cases:
	\begin{enumerate}
		\item If at least one of the speeds $\omega$ or $a$ is different from zero for the design in (\ref{eq: dvf}), and $h$ is sufficiently large in (\ref{eq: pmov}), then the matrix $hK\tilde L$ has one eigenvalue equals $0$ and one eigenvalue equals $-\tilde\kappa(a + \iota\omega)$ whose corresponding eigenvectors are $\mathbf{1}_n$ and $(\frac{v^*}{a + \iota\omega}\mathbf{1}_n + p^*)$ respectively with $v^* = (v_x^* + v_y^*)$ being the designed translational velocity in (\ref{eq: dvf}). The rest of eigenvalues are in the right-half complex plane.
		\item If $\omega = a = 0$ and $v^*\neq 0$. Then, the matrix $hK\tilde L$ has $0$ as an eigenvalue with algebraic multiplicity $2$ but geometric multiplicity equals $1$. Furthermore, the chain of generalized (complex) eigenvectors associated with the eigenvalue $0$ is $\{-\tilde\kappa v^* \mathbf{1}_n, p^*\}$. In addition, if $h$ is sufficiently large, then the rest of eigenvalues are in the right-half complex plane.
	\end{enumerate}
\end{lemma}

\begin{pf}
The original complex Laplacian matrix $L$ has two zeros as eigenvalues with algebraic and geometric multiplicity equal $2$ whose two independent eigenvectors are $\mathbf{1}_n$ and $p^*$ respectively, i.e., according to the kernel $\mathcal{S}$ of $L$ we have that $L\mathbf{1}_n = 0$ and $Lp^* = 0$, and we note that the zero eigenvalues of $hKL$ and their eigenvectors are the same as in $L$. 

	Let us analyze the first case. Since $B^T\mathbf{1}_n=0$, we have that $hK\tilde L\mathbf{1}_n = (hKL - \tilde\kappa M B^T)\mathbf{1}_n = 0$, and
\begin{align}
	&(hKL - \tilde\kappa M B^T) (\frac{v^*}{a + \iota\omega}\mathbf{1}_n + p^*) = \nonumber \\
	&= -\tilde\kappa (M_{t_x}B^T p^* - M_{t_y}B^T p^* - M_sB^T p^* - M_rB^Tp^*) \nonumber \\
	&= - \tilde\kappa v^* \mathbf{1}_n - \tilde\kappa(a + \iota\omega)p^* \nonumber \\
	&= - \tilde\kappa(a + \iota\omega)(\frac{v^*}{a + \iota\omega}\mathbf{1}_n + p^*).
\end{align}
For the rest of eigenvalues of $hK\tilde L$, we can get them arbitrarily close to the eigenvalues of $hKL$ by increasing the value of $h$ (see equation (\ref{eq: Ltilde})). Also, all the eigenvalues of $hKL$ (excepting the two zero eigenvalues that were \emph{relocated} in $hK\tilde L$) can be placed arbitrarily far from the imaginary axis because of $K$, then, we can conclude that $hK\tilde L$ has one eigenvalue equals $0$ whose eigenvector is $\mathbf{1}_n$, another eigenvalue equals $- \tilde\kappa(a + \iota\omega)$ whose eigenvector is $(\frac{v^*}{a + \iota\omega}\mathbf{1}_n + p^*)$, and the rest of eigenvalues are on the right-half complex plane.

	Now, let us analyze the second case. For $hK\tilde L$ with $\omega = a = 0$, we have that
\begin{align}
	hK\tilde Lp^* = (hKL - \tilde\kappa (M_{t_x} + M_{t_y})B^T)p^* = -\tilde\kappa v^* \mathbf{1}_n,
	\label{eq: genv}
\end{align}
and
\begin{align}
	&hK\tilde L (-\tilde\kappa v^*\mathbf{1}_n) = -h\tilde\kappa v^* K\tilde L \mathbf{1}_n = 0,
	\label{eq: genv2}
\end{align}
	consequently, we have that $(hK\tilde L)^2p^* = 0$ but $hK\tilde Lp^* \neq 0$, and $hK\tilde L \mathbf{1}_n = 0$. Now we check the algebraic multiplicity of the zero eigenvalue of $hK\tilde L$. Again, the rest of eigenvalues of $hK\tilde L$ can be placed arbitrarily closed to $hKL$ by increasing $h$. Besides, the non-zero eigenvalues of $hKL$ can stay \emph{far} from zero as much as we want because of $K$. Therefore, we can deduce then that from (\ref{eq: genv}) and (\ref{eq: genv2}) the matrix $hK\tilde L$ must have an eigenvalue $0$ with algebraic multiplicity $2$ (as $hKL$) but geometric multiplicity $1$, whose chain of generalized (complex) eigenvectors is $\{-\tilde\kappa v^*\mathbf{1}_n, p^*\}$, and the rest of eigenvalues are far from zero on the right-half complex plane. \qed
\end{pf}

Now we are ready for the main result. We remind that we will calculate a lower bound for the gain $h$ in Subsection \ref{sec: kappa}.

\begin{theorem}
\label{pro: comt}
Given a framework $\mathcal{F} = (\mathcal{G},p)$, whose graph is \textit{2-rooted} so that there exist weights $\omega_{ij}$ that satisfy the constrains (\ref{eq: wcomcon}) for a desired shape $\mathcal{S}$ as in (\ref{eq: dshape}) constructed from the reference shape $p^*$ as in (\ref{eq: pstar}). Consider the following control law for the dynamics (\ref{eq: dyn})
\begin{equation}
	u_i = -hk_i \sum_{j\in\mathcal{N}_i} \tilde\omega_{ij}(p_i - p_j), \quad \forall i\in\mathcal{V},
	\label{eq: uimodcomt}
\end{equation}
	where $h\in\mathbb{R}^+$ is sufficiently large as required in Lemma \ref{lem: eigtr}, $\tilde \omega_{ij}$ is as in (\ref{eq: wmod}), and $k_i$ has been designed such that the matrix $hKL$ does not have eigenvalues on the left-half complex plane. Consider the following two cases:
	\begin{enumerate}
		\item If $a=\omega=0$ and ${^b}v^* = ({^b}v_x^* + {^b}v_y^*) \neq 0$  in (\ref{eq: dvf}), then $p(t)\to \mathcal{S}$, where $\dot p_i(t) \to -c_2\tilde\kappa \, v^*, \forall i\in\mathcal{V}$, as $t\to\infty$, and $c_2 \in \mathbb{C}$, which depends on the initial condition $p(0)$, determines the eventual fixed scale of the formation with respect to $p^*$, and the orientation of $O_b$ with respect to $O_g$.
\item If at least one of the desired speeds $\omega$ or $a$ in (\ref{eq: dvf}) is different from zero, then $p(t)\to \mathcal{S}$ as $t\to\infty$. This eventual configuration $p(t)\in\mathcal{S}$ will describe a motion compatible with $\mathcal{S}$, i.e., $p(t)\to c_1\mathbf{1}_n + c_2(\frac{v^*}{a + \iota\omega}\mathbf{1}_n + p^*)e^{\tilde\kappa(a + \iota\omega) t}$ as $t\to\infty$, where $c_{\{1,2\}}\in\mathbb{C}$ depend on the initial condition $p(0)$.
\end{enumerate}
\end{theorem}
\begin{pf}
	Let us consider the first case. According to Lemma \ref{lem: eigtr}, the non-zero eigenvalues of $-hK\tilde L$ are in the half-left complex plane if $h > 0$ is sufficiently large. Furthermore, the eigenvalue $0$ of $-hK\tilde L$ has algebraic multiplicity $2$ but geometric multiplicity $1$, and its chain of generalized eigenvectors is $\{-\tilde\kappa v^*\mathbf{1}_n,p^*\}$. Therefore, from the exponential of the Jordan form of ($-hK\tilde L$) we have that the solution of the closed-loop system (\ref{eq: pmov}) derived from (\ref{eq: uimodcomt}) is
\begin{equation}
	p(t) = -c_1 \tilde\kappa v^*\mathbf{1}_n + c_2 (p^* - \tilde\kappa v^*\mathbf{1}_n t) + \sum_{l=3}^n f_le^{\lambda_l t},
\label{eq: solptcom}
\end{equation}
	where $c_1,c_2\in\mathbb{C}$ depend on the initial condition $p(0)$. For the rest of eigenvalues of $-hK\tilde L$ we have that $\operatorname{Re}\{\lambda_l\} < 0,  l\geq 3$, and $f_l(t,c_l,w_l,w_{l-1},\dots,w_{l-g})$ are different functions, with $g\in\mathbb{N}$ depending on the algebraic and geometric multiplicity of $\lambda_i$, corresponding to linear combinations like $c_l(w_l + w_{l-1}t +w_{l-2}\frac{t^2}{2!}+\dots+w_{l-g}\frac{t^g}{g!})$ depending on the (possibly generalized) eigenvectors $w_l$ of $-hK\tilde L$ and constants $c_l$ given by the initial condition $p(0)$. Nevertheless, since these functions $f_l$ (polynomials on $t$) are multiplied by exponentials $e^{\lambda t}$, when we focus on the limit of (\ref{eq: solptcom}) for $t\to\infty$, we have that $p(t) \to -\tilde\kappa v^* \mathbf{1}_n(c_1 + c_2t) + c_2p^*$
	and $\dot p(t) \to -c_2 \tilde\kappa v^*\mathbf{1}_n$. Since ${^b}v^*$ is a designed complex constant, then we can conclude that $p(t) \to \mathcal{S}$ as $t\to\infty$. Furthermore, we can conclude that $c_2$ will determine the eventual direction and speed of the velocity of the formation in global coordinates with origin at $O_g$. In particular, the speed will be proportional to the scale of $p(t)\in\mathcal{S}$ with respect to $p^*$, which is given by $c_2$ as well. Consequently, the eventual configuration $^bp(t)$ will travel with the constant translational velocity $|c_2|\tilde\kappa \, ^bv^*\mathbf{1}_n$ whose \emph{heading} or orientation of $O_b$ with respect to $O_g$ will depend on $c_2$. 

	Now, let us consider the second case. Similarly as in the first case, if $h > 0$ is sufficiently large, then, according to Lemma \ref{lem: eigtr}, the solution of the closed loop (\ref{eq: pmov}) derived from (\ref{eq: uimodcomt}) is given by
\begin{equation}
	p(t) = c_1 \mathbf{1}_n + c_2(\frac{v^*}{a + \iota\omega}\mathbf{1}_n + p^*)e^{\tilde\kappa(a + \iota\omega) t} + \sum_{l=3}^n f_le^{\lambda_l t}, \nonumber
\end{equation}
	where $c_1,c_2\in\mathbb{C}$ depend on the initial condition $p(0)$, the eigenvalues of $-hK\tilde L$ satisfy $\operatorname{Re}\{\lambda_l\} < 0,  l\geq 3$, and $f_l$ are polynomial functions on $t$ as discussed in the first case. Then, we can deduce that $p(t)\to\mathcal{S}$ as $t\to\infty$, since $p(t)\to c_1\mathbf{1}_n +  c_2(\frac{v^*}{a + \iota\omega}\mathbf{1}_n + p^*)e^{\tilde\kappa(a + \iota\omega) t}$ as $t\to\infty$.\qed
\end{pf}

\subsection{A lower bound for the gain $h$}
\label{sec: kappa}
Our strategy to find an lower bound for $h$ focuses on showing that $p(t) \to \operatorname{Ker}\{L\}$, i.e., $p(t)\to\mathcal{S}$ as $t\to\infty$, via one Lyapunov analysis so that $h$ can be taken into account explicitly in the stability analysis. Note that the kernel of $L$ and $hKL$ is the same. Let us define the following complementary subspace $\mathcal{S}^\perp := (\operatorname{Ker}\{L\})^\perp$ and let $P_\mathcal{S}, P_{\mathcal{S}^\perp}\in\mathbb{C}^{n\times n}$ be the projection matrices over the spaces $\mathcal{S}$ and $\mathcal{S}^\perp$ respectively. We then split
\begin{equation}
	p = P_\mathcal{S}\, p + P_{\mathcal{S}^\perp}\, p = p_\parallel + p_\perp.
\end{equation}
We are interested in showing that $p_\perp(t) \to 0$ as $t\to\infty$. Let us write the following dynamics derived from (\ref{eq: pmov})
\begin{align}
	\dot p_\perp &=  P_{\mathcal{S}^\perp} \dot p =  -P_{\mathcal{S}^\perp} hKL(p_\parallel + p_\perp) + P_{\mathcal{S}^\perp}\tilde\kappa MB^T(p_\parallel + p_\perp).
	\label{eq: aux76}
\end{align}
We have shown in Lemma \ref{lem: eigtr} that $MB^Tp_\parallel\in\mathcal{S}$, i.e., $P_{\mathcal{S}^\perp} MB^Tp_\parallel = 0$, and together with $KL p_\parallel = 0$ and $P_{\mathcal{S}^\perp}KL p_\perp = KL p_\perp$, we can simplify (\ref{eq: aux76}) as
\begin{equation}
	\dot p_\perp = -hKLp_\perp + \tilde\kappa P_{\mathcal{S}^\perp} MB^T p_\perp.
	\label{eq: aux77}
\end{equation}

Consider the Jordan form $J\in\mathbb{C}^{n\times n}$ of $KL$, i.e., $J = TKLT^{-1} = \left[\begin{smallmatrix}J_1 & 0 \\ 0 & J_2 \end{smallmatrix}\right]$ for some invertible matrix $T$, and $J_1\in\mathbb{C}^{2\times 2}$ and $J_2\in\mathbb{C}^{n-2\times n-2}$. In fact, let the first submatrix $J_1$ corresponds to the two zero eigenvalues of $KL$, i.e., the zero matrix since each zero eigenvalue has an independent eigenvector ($\mathbf{1}_n$ and $p^*$). Consequently, $-J_2$ is a Hurwitz matrix. Now consider the coordinate transformation $Tp = \begin{bmatrix}q_1^T & q_2^T\end{bmatrix}^T,$
with $q_1\in\mathbb{C}^2$ and $q_2\in\mathbb{C}^{n-2}$. Since $J_1$ corresponds to the zero eigenvalues of $KL$, then we have that $Tp_\parallel = \left[\begin{smallmatrix}q_1^T & 0\end{smallmatrix}\right]^T$ and $Tp_\perp = \left[\begin{smallmatrix}0 & q_2^T\end{smallmatrix}\right]^T$. Hence, by applying the same coordinate transformation to the dynamics (\ref{eq: aux77}) we have that
\begin{align}
	&\frac{\mathrm{d}}{\mathrm{dt}}\left[\begin{smallmatrix}0 \\ q_2\end{smallmatrix}\right] = -hTKLT^{-1}\left[\begin{smallmatrix}0 \\ q_2\end{smallmatrix}\right] + \tilde\kappa TP_{\mathcal{S}^\perp}  MB^TT^{-1}\left[\begin{smallmatrix}0 \\ q_2\end{smallmatrix}\right] \nonumber \\
		\dot q_2 &= -hJ_2 q_2 + \tilde\kappa \left(TP_{\mathcal{S}^\perp} MB^TT^{-1}\right)_{\scriptscriptstyle (n-2 \times n-2)}q_2,
	\label{eq: aux79}
\end{align}
where the subindex for the matrix in the second term of (\ref{eq: aux79}) means that we have eliminated its first two rows and two columns. Then if $q_2(t)$ converges asymptotically to zero, then $p_\perp(t)$ will do it as well. Given a fixed $M$ and $\tilde\kappa$ that were designed in Section \ref{sec: mpd} for a desired collective motion without the need of $h$, we can calculate a lower bound for $h$ that guarantees the convergence to the origin of the linear system (\ref{eq: aux79}) with a standard Lyapunov analysis.
\begin{proposition}
\label{pro: kappa}
	The system (\ref{eq: aux79}) is exponentially stable if $h > \tilde\kappa ||Q\left(MB^T\right)_{\scriptscriptstyle(n-2 \times n-2)}||_2$, where $Q$ is a Hermitian positive definite matrix such that $QJ_2 + J_2^HQ \geq 2I_{(n-2)}$, where $X^H$ denotes the conjugate transpose of $X$.
\end{proposition}
\begin{pf}
Since $-J_2$ is Hurwitz matrix, then there exists a Hermitian positive definite matrix $Q$ such that $QJ_2 + J_2^HQ \geq 2I_{(n-2)}$. Now consider the Lyapunov candidate $V = q_2^HQq_2$, whose time derivative satisfies
{\small
	$
	\frac{\mathrm{d}V}{\mathrm{dt}} \leq -2h||q_2||^2 +2\tilde\kappa \, ||Q\left(TP_{\mathcal{S}^\perp} MB^TT^{-1}\right)_{\scriptscriptstyle (n-2 \times n-2)}||_2 \, ||q_2||^2$.
	
}
	We note that $||P_{\mathcal{S}^\perp}||_2 = 1$, and that we do and undo a change of coordinates with $T$; thus the system (\ref{eq: aux79}) is stable if $h > \tilde\kappa ||Q\left(MB^T\right)_{\scriptscriptstyle (n-2 \times n-2)}||_2.$
\qed
\end{pf}
The following algorithm summarizes the design process for the modified weights in (\ref{eq: wmod}) to guarantee the convergence of the formation to the desired eventual collective motion by implemeting the control law (\ref{eq: uimodcomt}).
\begin{alg}
\begin{enumerate}
\item Given a desired $p^*$, we calculate the weights $\omega_{ij}$ and the gains $k_i$ according to \cite{lin2014distributed} to construct the Laplacian $L$ and the gain matrix $K$ in (\ref{eq: comcom}).
\item We calculate the sets of motion parameters $\mu_{ij}$ for $M_{t_{\{x,y\}}}, M_s$, and $M_r$ in (\ref{eq: ormo}) for the horizontal/vertical translation, scaling, and rotation so that they correspond to a $1$ horizontal/vertical \emph{distance unit}/sec, $1$ \emph{current size}/sec scaling, and $1$ radian/sec respectively.
\item We finish the design of $M$ in (\ref{eq: ormo}) with choosing the coordinates $\kappa_{\{t_{\{x,y\}},s,r\}}$ that define the eventual collective motion of the formation. Then, we choose $\tilde\kappa$ to set the \emph{global speed} of the collective motion.
\item Finally, we calculate $h$ for the modified weights in (\ref{eq: wmod}) and the control action (\ref{eq: uimodcomt}) such that the condition in Proposition \ref{pro: kappa} is satisfied. 
\end{enumerate}
\end{alg}



\section{Applications, simulations and comparisons}
\label{sec: simulations}

\begin{figure}
\centering
\includegraphics[width=0.8\columnwidth]{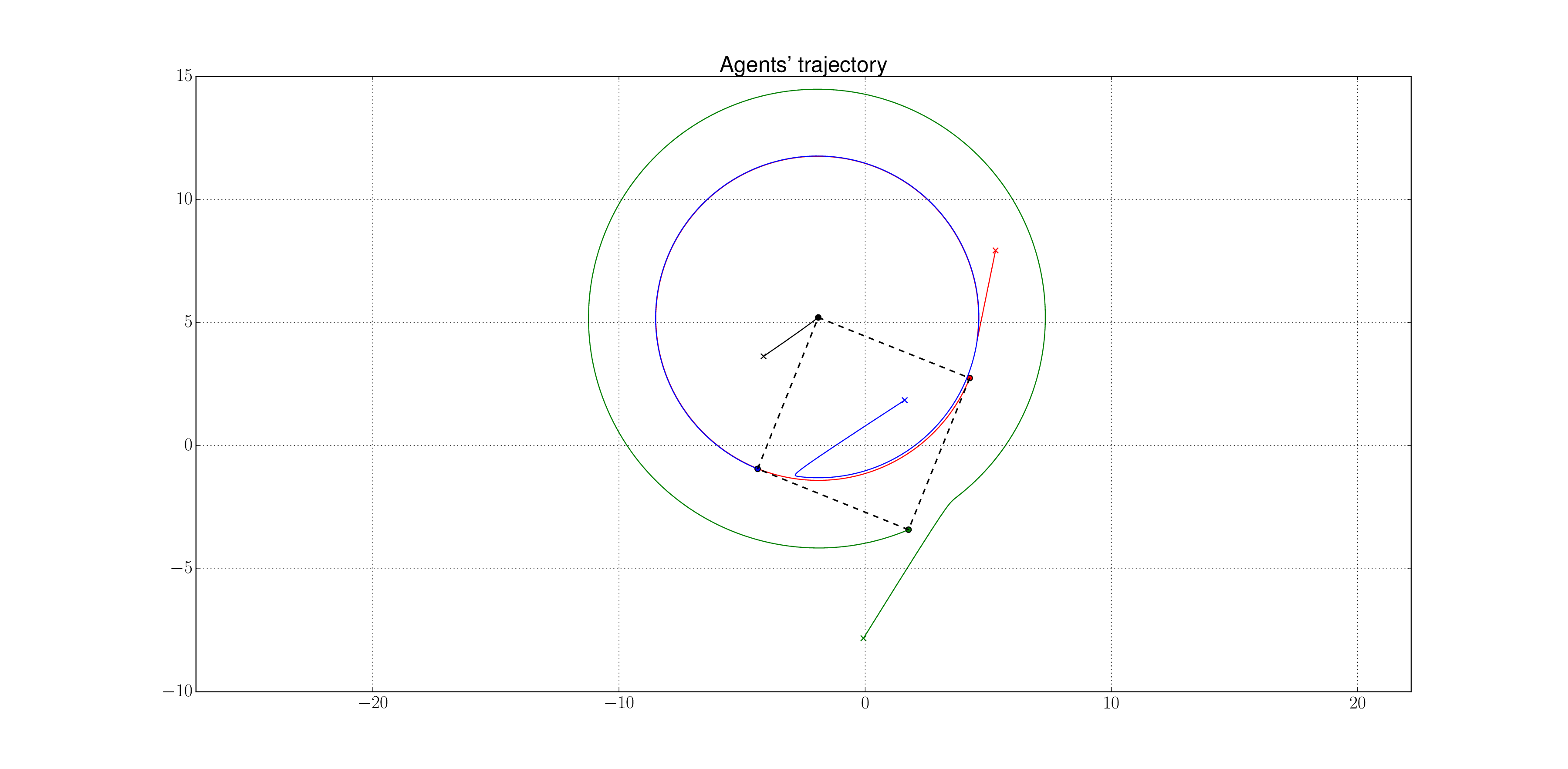}
	\caption{The formation converges to a square and cooperates to enclose and orbit around one of its agents (in black color). The desired rotational and translational velocities of the formation have been designed such that the enclosed agent is the instantaneous center of rotation. The crosses and the dots denote for the initial and the $t = 250$ secs positions. The dashed lines denote the edges of the graph.}
\label{fig: enc}
\end{figure}

\emph{Enclosing of a target}: Let us start with a team of four agents with the following square reference shape $p^* = \left[\begin{smallmatrix}0 & -\iota & (1-\iota) & 1\end{smallmatrix}\right]^T \in\mathbb{C}^4$, and the incidence matrix $B$ is constructed with the following ordered set of edges $\mathcal{Z}=\{(1,2),(2,3), (3,4), (4,1)\}$. We design $M$ such the agents $2,3$ and $4$ will describe a circular orbit around the agent $1$. We first design $M$ with the velocities as in Figure \ref{fig: velcom}
\begin{align} \scriptstyle
	M &= \scriptstyle\kappa_{t_x}
\left[\begin{smallmatrix}
	0 & 0 & 0 & 1 \\ 0 & -1 & 0 & 0 \\ 0 & -1 & 0 & 0 \\ 0 & 0 & 0 & 1
\end{smallmatrix}\right] +  \scriptstyle\kappa_{t_y}
	\left[\begin{smallmatrix}
	0 & 0 & 0 & -\iota \\ 0 & \iota & 0 & 0 \\ 0 & \iota & 0 & 0 \\ 0 & 0 & 0 & -\iota
	\end{smallmatrix}\right] + \scriptstyle\kappa_{r}
	\left[\begin{smallmatrix}
	0 & 0 & 0 & -1+\iota \\ 0 & -1-\iota & 0 & 0 \\ 0 & -1+\iota & 0 & 0 \\ 0 & 0 & 0 & -1-\iota
	\end{smallmatrix}\right] + \nonumber \\ &  \scriptstyle\kappa_{s}
	\left[\begin{smallmatrix}
	0 & 0 & 0 & -1-\iota \\ 0 & 1-\iota & 0 & 0 \\ 0 & -1-\iota & 0 & 0 \\ 0 & 0 & 0 & 1-\iota
	\end{smallmatrix}\right]. \nonumber
\end{align}
One can check that the following coordinates $\kappa_{\{t_{\{x,y\}},r\}} = 1$, together with $\kappa_s = 0$ for no scaling, achieves such a collective motion.
We set as a \emph{global speed} for the collective motion $\tilde\kappa = 1$. For the construction of $L$, we follow the design shown in \cite{Marina2017}, i.e., the weights $\omega_{ij}$ for the agent $i\in\mathcal{V}$ are $\omega_{i\,\,\operatorname{mod}_{n}(i-1)} = e^{-\iota \frac{\pi}{2}(1-\frac{n-2}{n})}$ and $\omega_{i\,\,\operatorname{mod}_{n}(i+1)} = e^{\iota \frac{\pi}{2}(1-\frac{n-2}{n})}$, with $n=4$, so that $K$ in (\ref{eq: comcom}) can be the identity matrix. In order to find the lower bound for $h$, we first calculate $J_2 = 2.82I_{2\times 2}$ resulting in $Q = 0.3536I_{2\times 2}$. Now we can calculate $||Q\left(MB^T\right)_{\scriptscriptstyle (n-2 \times n-2)}||_2 = 1$, and since $\tilde\kappa$ has been set to $1$, then we choose $h = 1.5$ in order to satisfy the condition in Proposition \ref{pro: kappa}. We show the simulation in Figure \ref{fig: enc}. A similar technique based on distance-based formation control has been proposed in \cite{Marina16}, where instead of weights the authors manipulate desired distances. While distance-based can guarantee collision avoidance between neighboring agents and can be employed in 3D (or higher), only local-convergence is given due to nonlinearities.

\begin{figure}
\centering
\includegraphics[width=0.48\columnwidth]{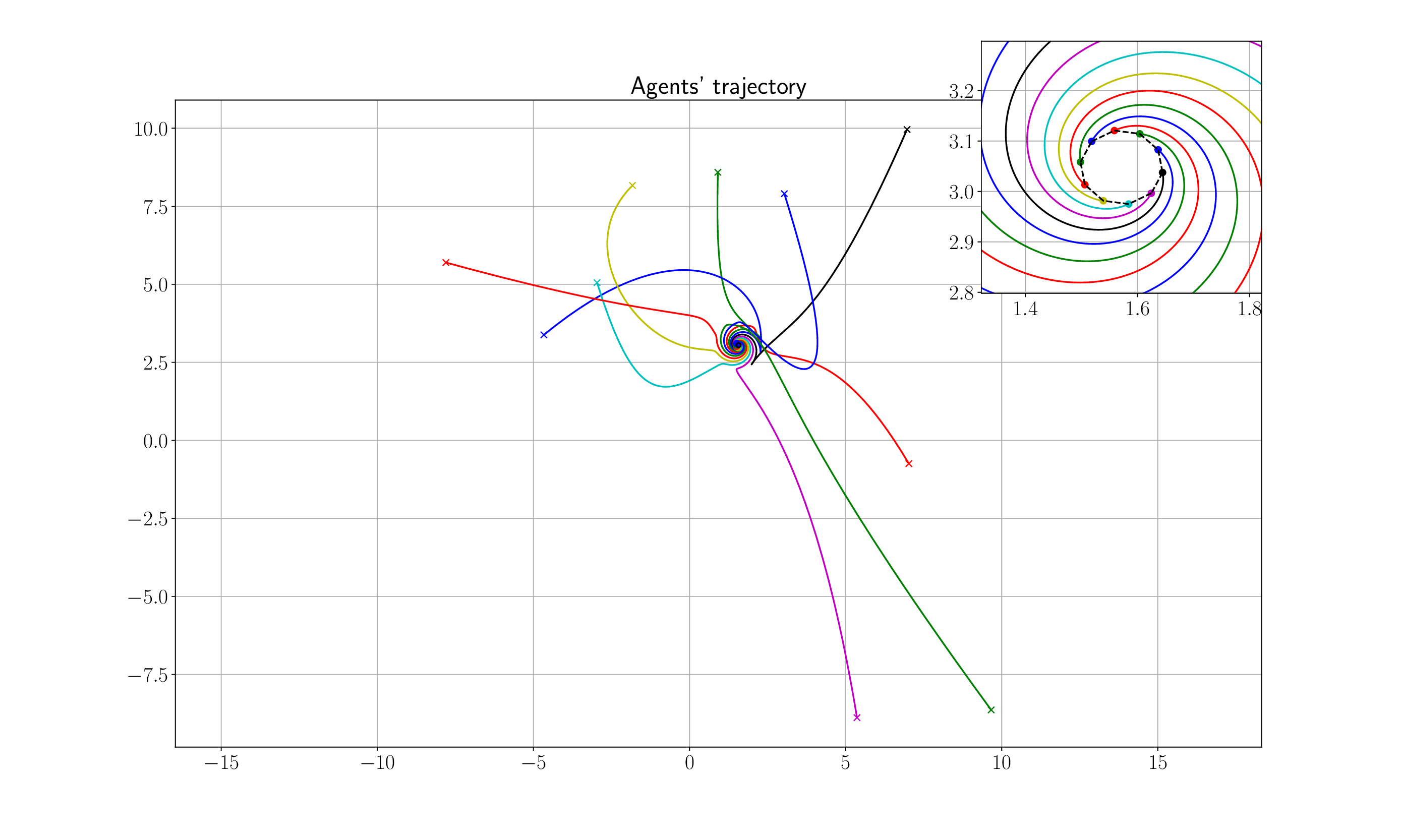}
\includegraphics[width=0.48\columnwidth]{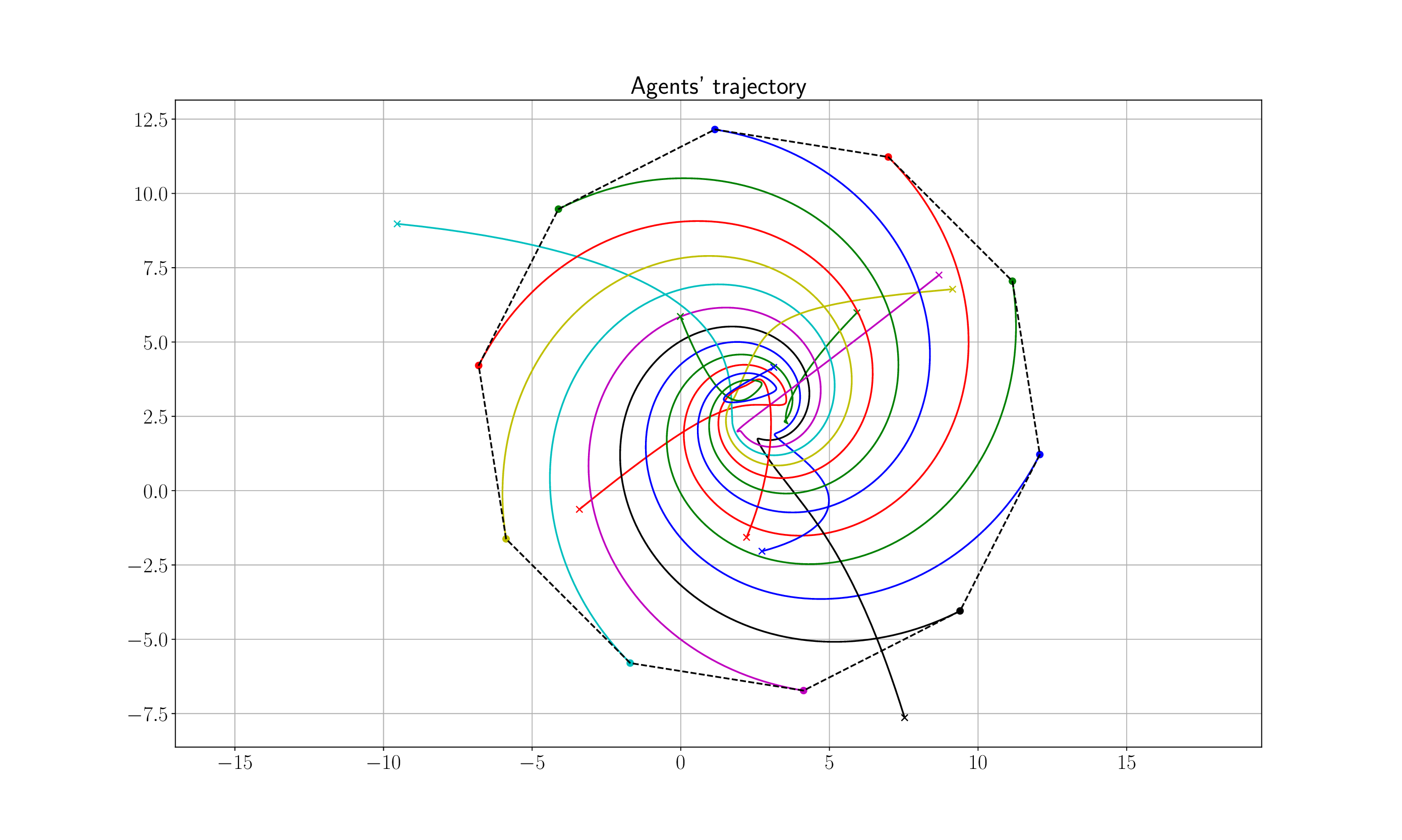}
	\caption{On the left-hand side, the formation eventually forms a regular decagon that describes a circular spiral towards its centroid, i.e., a \emph{shaped} consensus. On the right-hand side the formation describes a circular spiral while the eventual displayed decagon grows exponentially fast. The crosses and the dots denote for the initial and the $t = 250$ secs positions. The dashed lines denote the edges of the graph.}
\label{fig: decacon}
\end{figure}

\emph{Shaped consensus}: If we consider the coordinates $\kappa_{\{t_{\{x,y\}}\}} = 0$ and $k_{\{r,s\}} = 1$, then a generic formation will describe an eventual spinning motion around its centroid together with a scaling motion, i.e., we can cover an area by describing circular spirals within $\mathcal{S}$ if $\kappa_s>0$. Conversely, we have the consensus of the formation, i.e., $p(t)\to b\mathbf{1}_n, b\in\mathbb{C}$ as $t\to\infty$ if we set $\kappa_s<0$. Our extension from the \emph{standard} consensus \cite{olfati2007consensus} is that in Theorem \ref{pro: comt} we set $p^*$ as the eigenvector associated to the algebraic connectivity $\lambda_2 = \kappa_s + \iota\kappa_r$ of the Laplacian matrix. Therefore, a \emph{shaped consensus} occurs, i.e, the formation will display the desired formation while rendezvous. We show both maneuvers in Figure \ref{fig: decacon} for a decagon. The weights $\omega_{ij}$ for the regular decagon were chosen as in the \emph{enclosing of a target scenario} but with $n = 10$. It can be checked that for a regular polygon
	\begin{equation}
\scriptstyle
		M_r = \left[
\begin{smallmatrix}
	1 & 0 & 0 & \cdots & 0 & 1 \\
	1 & 1 & 0 & \cdots & 0 & 0 \\
	0 & 1 & 1 & \cdots & 0 & 0 \\
	\vdots & \vdots & \vdots & \vdots & \vdots & \vdots \\
		0 & 0 & 0 & \cdots & 1 & 1
\end{smallmatrix}\right], 
M_s = \left[
\begin{smallmatrix}
	1 & 0 & 0 & \cdots & 0 & -1 \\
	-1 & 1 & 0 & \cdots & 0 & 0 \\
	0 & -1 & 1 & \cdots & 0 & 0 \\
	\vdots & \vdots & \vdots & \vdots & \vdots & \vdots \\
		0 & 0 & 0 & \cdots & -1 & 1
\end{smallmatrix}\right].
\nonumber
\end{equation}

\emph{Translation with controlled speed and heading}: Let us consider the coordinates $\kappa_{\{t_y,s,r\}} = 0$ and $\kappa_{t_x} = 1$. Then, according to Theorem \ref{pro: comt}, the formation will travel with constant velocity eventually. However, the translational velocity has been designed with respect to $O_b$ and not $O_g$ as it is shown in Figure \ref{fig: velcom}. We could choose one agent $i\in\mathcal{V}$ to control one relative position $z_{ij}, j\in\mathcal{N}_i$, in global coordinates, e.g., with adding the simple proportional controller $-(z_{ij} - z_{ij}^*)$ to its control action with $z_{ij}^*\in\mathbb{C}$ with respect to $O_g$ and compatible with $\mathcal{S}$. Since $z_{ij}^*$ has both direction and magnitude, then we are controlling both, the heading of the translational velocity in $O_g$ and its speed since it depends on the size of $p(t)\in\mathcal{S}$. This approach is free of any extra complexities/estimators as it is required in a \emph{leader-follower} approach \cite{lin2013leader,zhao2018affine}.

\begin{figure}
\centering
\includegraphics[width=1\columnwidth]{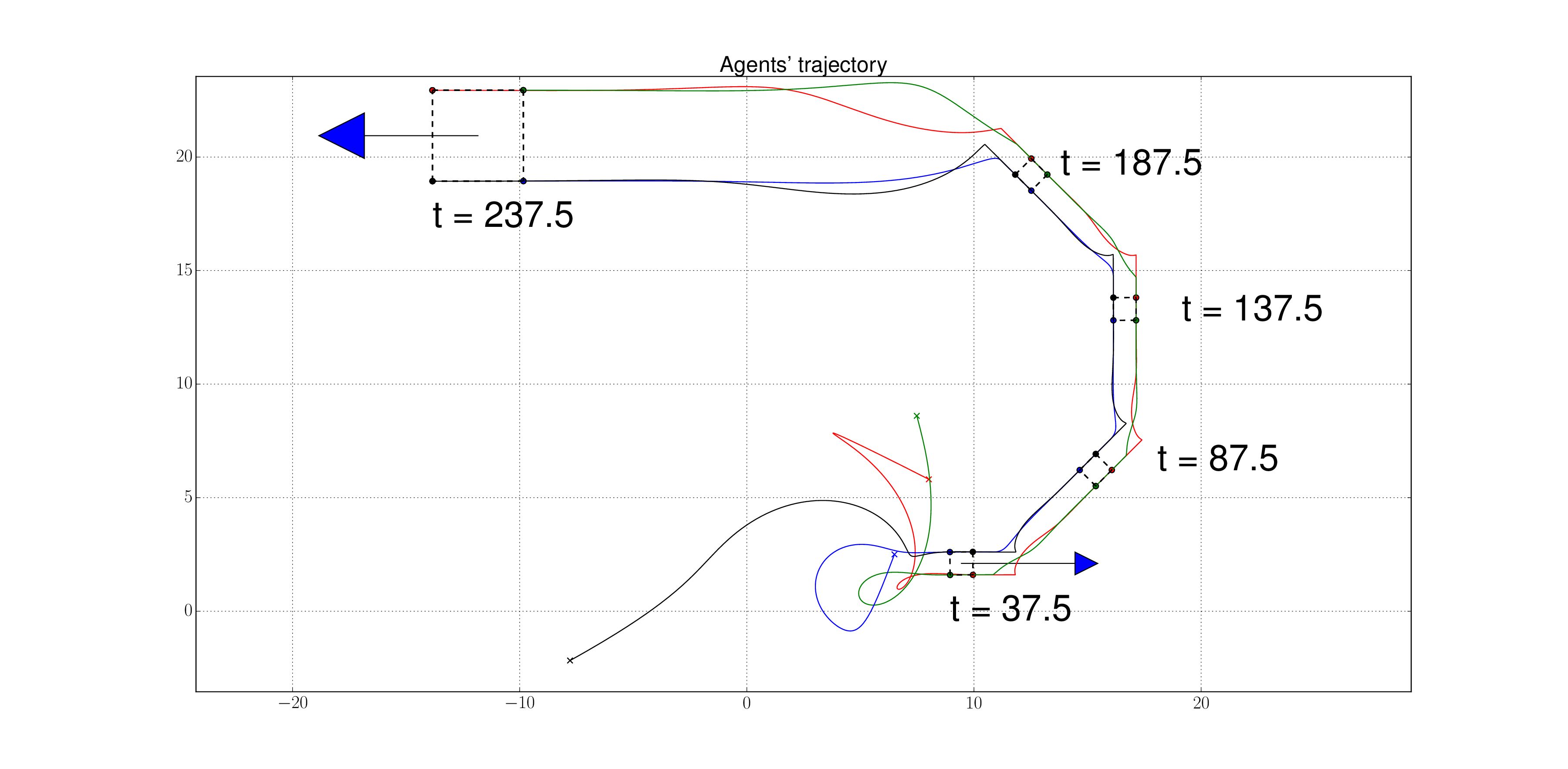}
	\caption{The formation converges to a square with only translational velocity w.r.t. $O_b$. The eventual formation's velocity (blue arrow) is designed to be parallel to the side $z_{12}$. The red agent controls the global orientation of such a side, e.g., with the proportional controller ($z_{12} - z_{12}^*$). Every $50$ seconds, the desired $z_{12}^*$ is rotated by $\frac{\pi}{4}$ radians. In the last rotation, $z_{12}^*$ is also four times bigger than at the beginning, hence the square not only grows, but the velocity's speed is increased accordingly. The crosses denote for the initial positions, and the dashed lines denote the edges of the graph.}
\label{fig: trav}
\end{figure}

\section{Conclusions}
\label{sec: conc}
We have presented a maneuvering technique to induce collective motions with global convergence in \emph{complex-Laplacian-based} formation control. This technique can be exploited in the problems of \emph{shape consensus}, \emph{enclosing of a target}, and \emph{travelling formation with controlled speed and heading}. 